\documentstyle[12pt,citesort]{elsart}
 
\newcommand{\ra}{\rangle}
\newcommand{\la}{\langle}
\begin{document}
\begin{frontmatter}

\title{$\pi N$ scattering and electromagnetic corrections \\
in the perturbative chiral quark model} 

\author[Tuebingen]{V.E. Lyubovitskij}, 
\author[Tuebingen]{Th. Gutsche}, 
\author[Tuebingen]{Amand Faessler}, 
\author[Paris]{R. Vinh Mau}
\address[Tuebingen]{Institut f\"ur Theoretische Physik, Universit\"at 
T\"ubingen, Auf der Morgenstelle 14, D-72076 T\"ubingen, Germany}
\address[Paris]{Laboratoire de Physique Th\'eorique des Particules 
\'El\'ementaires, Universit\'e P. et M. Curie, 4 Place Jussieu, 
75252 Paris Cedex 05, France}
 
\date{\today}
 
\maketitle

\vskip.5cm

\begin{abstract}
We apply the perturbative chiral quark model to give predictions for the 
electromagnetic $O(p^2)$ low-energy couplings of the ChPT effective 
Lagrangian that define the electromagnetic mass shifts of nucleons and 
first-order $(e^2)$ radiative corrections to the $\pi N$ scattering amplitude. 
We estimate the leading isospin-breaking correction to the strong energy 
shift of the $\pi^- p$ atom in the $1s$ state, which is relevant for the 
experiment "Pionic Hydrogen" at PSI. 

\vskip .3cm

\noindent {\it PACS:} 
11.10.Ef, 12.39.Fe, 12.39.Ki, 13.40.Dk, 13.40.Ks, 14.20.Dh 
       
\noindent {\it Keywords:} Relativistic chiral quark model; 
Pion-nucleon amplitude; Radiative corrections; Hadronic atoms.  

\end{abstract}

\end{frontmatter}
In \cite{Weinberg1,Tomozawa} Weinberg and Tomozawa derived a model-independent 
expression for the $S$ -wave $\pi N$ scattering lengths using the current 
algebra relations and the PCAC assumption. To reproduce the result for the 
$\pi N$ scattering lengths one can use the specific Lagrangian with the 
nucleon field $N$ refered to as the Weinberg-Tomozawa (WT) term  
\cite{Weinberg2}-\cite{Leutwyler1} which is part of the effective Weinberg 
Lagrangian. The effective Weinberg Lagrangian can be derived from the original 
$\sigma$-model \cite{Gell-Mann_Levy} by performing a chiral-field dependent 
rotation on the nucleon field \cite{Weinberg2}. On the quark level 
the same exercise was done in the framework of the cloudy bag 
model \cite{Thomas1,Jennings}. The chiral transformation 
eliminates the nonderivative coupling of the chiral (pion) field with the 
nucleons/quarks and replaces it by a nonlinear derivative coupling (axial 
vector term + WT term + higher order terms in the chiral field). Note, that 
both realizations of chirally-symmetric Lagrangians (the original 
$\sigma$-model and the Weinberg type Lagrangian) should \'{a} priori give the 
same result for the $\pi N$ $S$-wave scattering lengths. 
In ref. \cite{Long_paper} in the framework of perturbative chiral quark model 
(PCQM) \cite{Gutsche,PCQM} we demonstrate that the equivalence between the two 
theories with nonderivative and derivative coupling of the chiral field to the 
quarks is also valid when including the photon field. 

The purpose of this letter is to calculate first-order ($e^2$) radiative 
corrections to the nucleon mass and the pion-nucleon amplitude at threshold. 
We thereby predict the $O(p^2)$ electromagnetic (e.m.) low-energy couplings 
(LECs) originally defined in the effective Lagrangian of Chiral Perturbation 
Theory (ChPT)~\cite{Meissner1,Meissner2}. Quantitative information about 
these constants is important for the ongoing experimental and theoretical 
analysis of decay properties of the $\pi^-p$ atom (for a detailed discussion 
see Ref. \cite{piP-atom}). In particular, we give a prediction for the 
leading isospin-breaking correction to the strong energy shift of the 
$\pi^- p$ atom in the $1s$ state. 
 
Following considerations are based on the perturbative chiral quark model 
(PCQM), a relativistic quark model suggested in \cite{Gutsche} and extended 
in \cite{PCQM} for the study of low-energy properties of baryons. The model 
includes relativistic quark wave functions and confinement as well as the 
chiral symmetry requirements. The quarks move in a self-consistent field, 
represented by scalar $S(r)$ and vector $V(r)$ components of a static 
potential with $r=|\vec{x}|$ providing confinement. The interaction of quarks 
with Goldstone bosons is introduced on the basis of the nonlinear 
$\sigma$-model \cite{Gell-Mann_Levy}. The PCQM is based on the effective, 
chirally invariant Lagrangian ${\cal L}_{inv}$ \cite{PCQM}  
\begin{eqnarray}\label{Linv} 
{\cal L}_{inv}(x)&=&\bar\psi(x) \biggl\{ i\not\! \partial -\gamma^0 V(r)
-  \, S(r) \, \biggl[ \frac{U \, + U^\dagger}{2} \, 
+ \, \gamma^5 \, \frac{U \, -  U^\dagger}{2} \biggr] \biggr\} \, \psi(x)\\
&+&\frac{F^2}{4} \, {\rm Tr} [\, \partial_\mu U \, 
\partial^\mu U^\dagger \, ]\nonumber  
\end{eqnarray}
where $\psi$ is the quark field, $U=\exp[i\hat\Phi/F]$ is the chiral field 
and $F=88$ MeV is the pion decay constant in the chiral 
limit \cite{Gasser1,PCQM}. In the following we restrict to the $SU(2)$  flavor 
case, that is  $\hat\Phi \to \hat\pi  = \vec{\pi} \vec{\tau}$. For small 
fluctuations of the mesons fields one can use the perturbation expansion in 
powers of the parameter $1/F$. The PCQM was successfully applied to 
$\sigma$-term physics and extended to the study of electromagnetic properties  
of the nucleon \cite{PCQM}. Similar perturbative quark models have also been 
studied in Refs. \cite{Thomas-Chin}. 

The quark field $\psi$ we expand in the basis of potential eigenstates as 
\begin{eqnarray}\label{total_psi} 
\psi(x)&=&\sum\limits_\alpha b_\alpha u_\alpha(\vec{x})
\exp(-i{\cal E}_\alpha t) - \sum\limits_\beta d_\beta^\dagger  
v_\beta(\vec{x}) \exp(i{\cal E}_\beta t)  
\end{eqnarray}
where the sets of quark $\{ u_\alpha \}$ and antiquark $\{ v_\beta \}$ wave 
functions in orbits $\alpha$ and $\beta$ are solutions of the Dirac equation 
with the static potential. The expansion coefficients $b_\alpha$ and 
$d_\beta^\dagger$ are the corresponding single quark annihilation and 
antiquark creation operators. 

The direct way to generate the WT term in the Lagrangian (\ref{Linv}) is 
through introduction of a unitary transformation on the quark field $\psi$. 
The technique was, for example, performed in the context of the cloudy bag 
model \cite{Thomas1}. With the unitary chiral rotation 
$\psi \to \exp\{ -i\gamma^5\hat{\Phi}/(2F) \} \psi$ the Lagrangian 
(\ref{Linv}) transforms into a Weinberg-type form ${\cal L}^W$ containing 
the axial-vector coupling and the WT term: 
\begin{eqnarray}\label{L_W}  
\hspace*{-.75cm}{\cal L}^W(x) &=& {\cal L}_0(x) \, + \, 
{\cal L}^{W; str}_I(x) \, + \, o(\vec{\pi}^2) ,\\ 
\hspace*{-.75cm}{\cal L}_0(x) &=& \bar\psi(x) \biggl\{ i\not\! \partial - S(r) 
- \gamma^0 V(r) \biggr\} \psi(x) 
- \frac{1}{2} \, \vec{\pi}(x) \, (\Box + M_\pi^2) \,\vec{\pi}(x) , \nonumber\\ 
\hspace*{-.75cm}
{\cal L}^{W; str}_I(x) &=& \frac{1}{2F} \partial_\mu \vec{\pi}(x) \, 
\bar\psi(x) \, \gamma^\mu \, \gamma^5 \, \vec{\tau} \, \psi(x) 
- \frac{\varepsilon_{ijk}}{4F^2} \, \pi_i(x) \, \partial_\mu \pi_j(x) \, 
\bar\psi(x) \, \gamma^\mu \, \tau_k \,  \psi(x) , \nonumber
\end{eqnarray}
where ${\cal L}^{W; str}_I$ is the $O(\pi^2)$ strong interaction Lagrangian, 
$\Box = \partial^\mu \partial_\mu$ and $M_\pi$ is the pion mass.

In ref. \cite{Long_paper} we demonstrate explicitly for the $\pi N$ amplitude 
up to order $(1/F^2)$ that the two effective theories, the original one 
involving the pseudoscalar coupling and the Weinberg type, are formally 
equivalent, both on the level of the Lagrangians and for the matrix elements. 
This equivalence is based on the unitary transformation of the quark fields, 
where, in addition, the quarks remain on their energy shell. The same relation 
also holds in a fully covariant formalism, when quarks/baryons are on their 
mass shell. Particularly, we show that the Weinberg-Tomozawa result can be 
reproduced with the use of the original Lagrangian (\ref{Linv}) if: i) we use 
the expansion of the chiral field up to quadratic terms and ii) we employ the 
full quark propagator including the antiquark components. The two forms of 
the Lagrangian also yield the same results when including the 
photon field. For the equivalence to hold it is essential that the photons are 
introduced consistently in both formalisms, that is by minimal substitution. 
One can prove that both Lagrangians yield the same results for radiative 
corrections to the $\pi N$ scattering amplitude at threshold. 

In this letter we apply the developed formalism to study e.m. corrections 
of nucleon properties, such as the mass and the $\pi N$ scattering amplitude. 
We perform all calculations using the technically more convenient Lagrangian 
(\ref{L_W}). Introduction of the e.m. field $A_\mu$ is accomplished by minimal 
substitution into Eq. (\ref{L_W}): 
\begin{eqnarray} 
\hspace*{-.5cm}
\partial_\mu\psi \to D_\mu\psi = \partial_\mu\psi + i e Q A_\mu \psi,  
\hspace*{.5cm} 
\partial_\mu\pi_i \to D_\mu\pi_i = \partial_\mu\pi_i + 
e \varepsilon_{3ij} A_\mu \pi_j
\end{eqnarray}
where $Q$ is the quark charge matrix.
 
Following the Gell-Mann and Low theorem \cite{Gell-Mann_Low} the 
e.m. mass shift $\Delta m_N^{em}$ of the nucleon with respect to 
the three-quark ground state $|\phi_0>^N$ is 
\begin{eqnarray}\label{Energy_shift} 
\Delta m_N^{em} \doteq  \,\, {^N\la}\phi_0| \, - \frac{i}{2} \, 
\int \, \delta(x^0) \, d^4x \, \int \, d^4y \, 
T[{\cal L}^{em}(x){\cal L}^{em}(y)] \, |\phi_0{\ra^N_c} 
\end{eqnarray}
to order $e^2$ in the e.m. interaction. Subscript "$c$" in 
Eq. (\ref{Energy_shift}) refers to contributions from connected graphs only. 
Superscript "$N$" indicates that 
the matrix elements have to be projected onto the respective nucleon states. 
These nucleon states are conventionally set up by the product of single quark 
$SU(6)$ spin-flavor and $SU(3)_c$ color w.f. (see details in \cite{PCQM}), 
where the nonrelativistic single quark spin wave function is replaced by the 
relativistic ground state solution. 
With the quark-photon interaction defined by the Lagrangian 
\begin{eqnarray}
{\cal L}^{em}(x) = - e A_\mu \bar\psi(x) Q \gamma^\mu \psi(x) , 
\end{eqnarray} 
the e.m. mass shift $\Delta m_N^{em}$ is generated by two diagrams: 
one-body (Fig.1a) and two-body diagram (Fig.1b).  

The leading e.m. corrections (up to order $e^2/F^2$) to the $\pi N$ 
scattering amplitude at threshold are generated by the interaction Lagrangian 
\begin{eqnarray}\label{L_int_W}  
{\cal L}_I^{W}(x) = {\cal L}_I^{W; str}(x) + {\cal L}_I^{W; em}(x)
\end{eqnarray} 
where ${\cal L}_I^{W; str}$ is given in Eq. (\ref{L_W}) and the additional 
e.m. part ${\cal L}_I^{W; em}$ is given by 
\begin{eqnarray}
{\cal L}_I^{W; em}(x) &=& {\cal L}^{em}(x) \, + \, \frac{e}{4F^2} A_\mu(x) 
\bar \psi(x) \gamma^\mu [ \vec{\pi}^{\, 2}(x) \tau_3 - \vec{\pi}(x) 
\vec{\tau} \pi^0(x) ] \psi(x) \\
&-&  e A_\mu(x) \varepsilon_{3ij} \biggl[ \pi_i(x) \partial^\mu \pi_j(x) 
\, - \, \frac{\pi_j(x)}{2F} \bar\psi(x) \gamma^\mu \gamma^5 \tau_i \psi(x) 
\biggr] \nonumber
\end{eqnarray} 
The $\pi N$ amplitude in the presence of $O(e^2)$ radiative corrections 
is given by 
\begin{eqnarray}\label{Key_Eq}
{^N\la}\phi_0; \pi_j| \, \sum\limits_{n=1}^{4} \frac{i^n}{n!} 
\int  \, d^4x_1 \ldots \int  d^4x_n \, \, T[{\cal L}_{I}^W(x_1) \ldots 
{\cal L}_I^W(x_n) \, ] \, |\phi_0; \pi_i{\ra^N_c} . 
\end{eqnarray}
The diagrams for $O(e^2/F^2)$ radiative 
corrections to the $\pi N$ amplitude at threshold are shown in Fig.2. 
To evaluate the diagrams in Figs.1 and 2 we use the photon propagator 
$D_{\mu\nu}$ in the Coulomb gauge\footnote{It can be shown that the results 
do not depend on the choice of the gauge.} to separate the contributions from 
Coulomb and transverse photons. 

First, we analyze the e.m. mass shift of the nucleon. 
The contributions of diagrams Fig.1a and Fig.1b are given by 
\begin{eqnarray}\label{one_full}
\Delta m^{em; a}_N &=& e^2 \, \cdot \, {^N\la}\phi_0| \int d^4x \int d^4y \, 
\delta(x^0) \, D_{\mu\nu}(x-y) \\
&\times&\bar \psi_0(x) \, \gamma^\mu \, Q \, iG_\psi(x,y) \, 
\gamma^\nu \, Q \, \psi_0(y) \, |\phi_0{\ra^N} \nonumber\\
\Delta m^{em; b}_N &=& \frac{e^2}{2} \, \cdot \, {^N\la}\phi_0| \int d^4x 
\int d^4y \, \delta(x^0) \, D_{\mu\nu}(x-y) \nonumber\\
&\times&\bar \psi_0(x) \, \gamma^\mu \, Q \, \psi_0(x) \, \bar \psi_0(y) \, 
\gamma^\nu \, Q \, \psi_0(y) \, |\phi_0{\ra^N} . \nonumber
\end{eqnarray}
where $iG_\psi(x,y) = <0|T\{\psi(x)\bar\psi(y)\}|0>$ is 
the quark propagator in a binding potential. In the following we truncate the 
expansion of the quark propagator to the 
ground state eigen mode:  
\begin{eqnarray}\label{quark_propagator_ground}  
iG_\psi(x,y) \to iG_0(x,y) \doteq u_0(\vec{x}) \, \bar u_0(\vec{y}) \, 
e^{-i{\cal E}_\alpha(x_0-y_0)} \, \theta(x_0-y_0), 
\end{eqnarray}
that is we restrict the intermediate baryon states to $N$ and $\Delta$ 
configurations. Inclusion of excited baryon states will be subject of future 
investigations. With the use of approximation (\ref{quark_propagator_ground}) 
$\Delta m^{em; a}_N$ and $\Delta m^{em; b}_N$ reduce to 
\begin{eqnarray}\label{one_two} 
\hspace*{-.8cm}
\Delta m^{em; a}_N&=&\frac{e^2}{16\pi^3} \la N| \sum\limits_{i=1}^3 
(Q^2)^{(i)} |N\ra \int \frac{d^3q}{\vec{q}^{\,\, 2}} 
\biggl\{ [G_E^p(-\vec{q}^{\, 2})]^2 - \frac{\vec{q}^{\,\, 2}}{2m_N^2} 
[G_M^p(-\vec{q}^{\, 2})]^2 \biggr\} \nonumber\\
\hspace*{-.8cm}
\Delta m^{em; b}_N&=&\frac{e^2}{16\pi^3}\int\frac{d^3 q}{\vec{q}^{\,\, 2}}\,\, 
\biggl\{ \la N|\sum\limits_{i\,\not\! = j}^3 
Q^{(i)}Q^{(j)}|N\ra [G_E^p(-\vec{q}^{\, 2})]^2 \\ 
\hspace*{-.8cm}
&-&\la N|\sum\limits_{i\,\not\! = j}^3 Q^{(i)} Q^{(j)} 
\vec{\sigma}^{(i)} \vec{\sigma}^{(j)}|N\ra \, \frac{\vec{q}^{\,\, 2}}{6m_N^2} 
[G_M^p(-\vec{q}^{\, 2})]^2 \biggr\} , \nonumber
\end{eqnarray}
where $|N \ra$ is the SU(6) spin-flavor w.f. of the nucleon. Here we introduce 
the proton charge $(G_E^p)$ and magnetic $(G_M^p)$ form factors (f.f.) 
calculated at zeroth order \cite{PCQM} (meson cloud corrections are not taken 
into account) with   
\begin{eqnarray}
& &\chi^\dagger_{N_{s^\prime}} \chi_{N_s} G_E(-\vec{q}^{\, 2}) = 
{^N\la}\phi_0|\int d^3x \, \bar \psi_0(\vec{x}) \, \gamma^0 \, 
\psi_0(\vec{x}) \, e^{ i \vec{q} \, \vec{x}}|\phi_0{\ra^N} ,\\ 
& & \chi^\dagger_{N_{s^\prime}} 
\frac{i \, [ \vec{\sigma}_N \times \vec{q} \, ]}{2m_N} 
\chi_{N_s} G_M(-\vec{q}^{\, 2}) = {^N\la}\phi_0|\int d^3x \, 
\bar \psi_0(\vec{x}) \, \vec{\gamma} \,\psi_0(\vec{x}) \, 
e^{i \vec{q} \, \vec{x}}|\phi_0{\ra^N} \nonumber
\end{eqnarray} 
where $\chi_{N_s}$ is the nucleon spin w.f. and  $\vec{\sigma}_N$ is the 
nucleon spin operator. Note that the contributions of Coulomb and transverse 
photons to the e.m. mass shifts (see Eqs. (\ref{one_two})) are 
related to the nucleon charge and magnetic f.f., respectively. The sum 
\begin{eqnarray}
\la N|\sum\limits_{i=1}^3 (Q^2)^{(i)}|N\ra   + 
\la N|\sum\limits_{i \, \not\! = j}^3 Q^{(i)}Q^{(j)}|N \ra = \left \{
\begin{array}{cc}
1   & \mbox{for} \,\,\, N=p\\
0   & \mbox{for} \,\,\, N=n\\
\end{array}
\right. 
\end{eqnarray}
is equivalent to the charge matrix of nucleons ($Q_N$ being the nucleon 
charge). In the limit $m_N \to \infty$ (when we neglect the contribution of 
$G_M^p$ in Eqs. (\ref{one_two})) we obtain for the e.m. mass shifts 
\begin{eqnarray}\label{Delta_mn}
\Delta m^{em}_N = \Delta m_N^{em; a} +  \Delta m_N^{em; b} =  
\frac{\alpha Q_N^2}{4\pi^2} 
\int \frac{d^3q}{\vec{q}^{\,\, 2}} \, [G_E^p(-\vec{q}^{\, 2})]^2  
\end{eqnarray}
consistent with the result (Eq. (12.4)) of Ref. \cite{Gasser_Leutwyler_PR}. 
Hence, the e.m. mass shift of the neutron vanishes in the heavy nucleon limit. 
  
In the numerical analysis we use the variational 
{\it Gaussian ansatz} \cite{PCQM} for the quark ground state wave function 
with the following analytical form: 
\begin{eqnarray}\label{Gaussian_Ansatz} 
u_0(\vec{x}) \, = \, N \, \exp\biggl[-\frac{\vec{x}^{\, 2}}{2R^2}\biggr] \, 
\left(
\begin{array}{c}
1\\
i \rho \, \vec{\sigma}\vec{x}/R\\
\end{array}
\right) 
\, \chi_s \, \chi_f \, \chi_c 
\end{eqnarray}      
where $N=[\pi^{3/2} R^3 (1+3\rho^2/2)]^{-1/2}$ is a constant fixed by the 
normalization condition $\int d^3x \, u^\dagger_0(x) \, u_0(x) \equiv 1$; 
$\chi_s$, $\chi_f$, $\chi_c$ are the spin, flavor and color quark wave 
functions, respectively. Our Gaussian ansatz contains two model parameters: 
the dimensional parameter $R$ and the dimensionless parameter $\rho$. 
The parameter $\rho$ can be related to the axial coupling constant $g_A$ 
calculated in zeroth-order (or the three quark-core) approximation: 
\begin{eqnarray}\label{ga_rho_match}
g_A=\frac{5}{3} \biggl(1 - \frac{2\rho^2} {1+\frac{3}{2} \rho^2} \biggr) . 
\end{eqnarray}
Therefore, $\rho$ can be replaced by the axial charge $g_A$ by means of the 
matching condition (\ref{ga_rho_match}).
The parameter $R$ can be physically understood as the mean radius of the 
three-quark core and  is related to the charge radius of the proton in the 
leading-order approximation as 
\begin{eqnarray}
\la r^2_E \ra^P_{LO} = \int d^3 x \, u^\dagger_0 (\vec{x}) \, 
\vec{x}^{\, 2} \, u_0(\vec{x}) \, = \, \frac{3R^2}{2} \, 
\frac{1 \, + \, \frac{5}{2} \, \rho^2}{1 \, + \, \frac{3}{2} \, \rho^2}. 
\end{eqnarray}
In our calculations we use the value $g_A$=1.25 obtained in ChPT 
\cite{Gasser1}. Therefore, we have only one free parameter, that is $R$. 
In the numerical studies \cite{PCQM} R is varied in the region from 0.55 fm to 
0.65 fm, which corresponds to a change of $\la r^2_E \ra^P_{LO}$  from 0.5 to 
0.7 fm$^2$. The exact Gaussian ansatz (\ref{Gaussian_Ansatz}) restricts the 
potentials $S(r)$ and $V(r)$ to a form proportional to $r^2$. They are 
expressed in terms of the parameters $R$ and $\rho$ (for details see 
Ref. \cite{PCQM}).  

Using (\ref{Gaussian_Ansatz}) the proton f.f. at zeroth order 
are determined as \cite{PCQM}: 
\begin{eqnarray}\label{EM_FF}
G_E(-\vec{q}^{\, 2}) &=& \exp\biggl(-\frac{\vec{q}^{\, 2} R^2}{4}\biggr) 
\biggl[ 1 - \frac{\vec{q}^{\, 2} R^2}{4} \kappa \biggr]  \\
G_M(-\vec{q}^{\, 2}) &=& 
\exp\biggl(-\frac{\vec{q}^{\, 2} R^2}{4}\biggr) 
2m_N R \, \sqrt{\kappa \biggl(1-\frac{3}{2}\kappa\biggr) }, 
\hspace*{.8cm} \kappa = \frac{1}{2} - \frac{3}{10}g_A . \nonumber 
\end{eqnarray}
With Eq. (\ref{EM_FF}) the e.m. mass shift is finally given as  
\begin{eqnarray}\label{Res_Delta_mn}
\Delta m^{em}_p &=& \frac{\alpha}{R\sqrt{2\pi}} 
\biggl[ 1 - \frac{\kappa}{2} + \frac{3}{16} \kappa^2  
- \frac{34}{9} \kappa \biggl(1 - \frac{3}{2} \kappa\biggr) \biggr] ,\\
\Delta m^{em}_n &=& - \frac{\alpha}{R\sqrt{2\pi}} 
\frac{8}{3} \kappa \biggl(1 - \frac{3}{2} \kappa\biggr) , \nonumber 
\end{eqnarray}
where $\alpha=1/137$ is the fine structure coupling. 
For our set of parameters $g_A=1.25$ and $R=0.6 \pm 0.05$ fm we get 
$\Delta m^{em}_p = 0.54 \pm 0.04$ MeV, $\Delta m^{em}_n = - 0.26 \pm 0.02$ MeV 
and $\Delta m^{em}_n - \Delta m^{em}_p = - 0.8 \pm 0.06$ MeV. These and the 
following uncertainties in our results correspond to the variation of the 
parameter $R$. Our predictions are in qualitative agreement with the results 
obtained by Gasser and Leutwyler using the Cottingham 
formula \cite{Gasser_Leutwyler_PR}: 
$\Delta m^{em}_p = 0.63$ MeV, $\Delta m^{em}_n = -0.13$ MeV,   
$\Delta m^{em}_n - \Delta m^{em}_p =  - 0.76$ MeV. 
To compare our prediction for the e.m. mass shifts of the nucleons 
with the result of ChPT \cite{Meissner2}, we recall the part of the ChPT 
Lagrangian \cite{Meissner2} which is responsible for radiative corrections 
\begin{eqnarray}\label{Lagrangian_ChPT} 
\hspace*{-.6cm}
{\cal L}_{ChPT}^{e^2} = e^2 \bar N \biggl\{ 
f_1 \biggl(1 - \frac{\vec{\pi}^{\, 2} - (\pi^0)^{\, 2}}{F^2} \biggr) \, 
+  \, \frac{f_2}{2} \biggl(\tau_3 - \frac{\vec{\pi}^{\, 2}\tau_3 - 
\pi^0 \vec{\pi} \vec{\tau}}{2F^2} \biggr)  + f_3 \biggr\}  N .  
\end{eqnarray}
The $O(p^2)$ low-energy constants (LECs) $f_1$, $f_2$ and $f_3$ contain 
the effect of the direct quark-photon interaction. Matching our results for 
the nucleon mass shifts to the predictions of ChPT \cite{Meissner2} with 
\begin{eqnarray}\label{Delta_m_CHPT}
\hspace*{-1cm} 
\Delta m^{em}_p|_{ChPT} = - 4\pi\alpha \biggl(f_1+f_3+\frac{f_2}{2}\biggr), 
\hspace*{.15cm}  
\Delta m^{em}_n|_{ChPT} = - 4\pi\alpha \biggl(f_1+f_3-\frac{f_2}{2}\biggr) 
\end{eqnarray}
we obtain following relations for the coupling constants $f_1$, $f_2$ and 
$f_3$: 
\begin{eqnarray}\label{f2_estimate} 
\hspace*{-.9cm} 
f_2 &=& - \frac{1}{2R (2\pi)^{3/2}} \biggl[ 1 - \frac{29}{18} \kappa 
+ \frac{89}{48} \kappa^2 \biggr] , \\ 
f_1 + f_3 &=& - \frac{1}{4R (2\pi)^{3/2}} \biggl[ 1 - \frac{125}{18} \kappa + 
\frac{473}{48} \kappa^2 \biggr] . \nonumber 
\end{eqnarray}
Our numerical result for $f_2 = - 8.7 \pm 0.7$ MeV is in good agreement with 
the value of $f_2 = - 8.3 \pm 3.3$ MeV \cite{Meissner2,piP-atom} extracted 
from the analysis of the elastic electron scattering cross section using the 
Cottingham formula \cite{Gasser_Leutwyler_PR}. For $f_1 + f_3$ we get 
$-1.5 \pm 0.1$ MeV.  

We furthermore give a prediction for the separate values of $f_1$, $f_3$ and 
the ratio $f_1/f_2$ as deduced from our model analysis of $e^2$ corrections to 
the $\pi N$ amplitude. We denote the corresponding matrix element associated 
with the nucleon flavor transition $N_1 \to N_2$ by $M_{N_1N_2}^{(e^2); ij}$. 
In the Coulomb gauge only six diagrams (Fig.2a-2f) contribute to the radiative 
correction to the $\pi N$ amplitude at threshold. The contribution of the 
other diagrams (Fig.2g-2o) vanishes. The contributions of the different 
diagrams of Fig.2 are as follow: 
\begin{eqnarray}\label{piN_Fig3ab} 
M_{N_1N_2}^{(e^2); ij}\bigg|_{a+b} &=& - \frac{e^2}{F^2} \, \cdot \, 
{^N\la}\phi_0| \int d^4x \int d^4y D_{\mu\nu}(x-y) \bar \psi_0(x) \gamma^\mu \\
&\times& ( T^{ij} G_\psi(x,y) Q + Q G_\psi(x,y) T^{ij}) 
\gamma^ \nu \psi_0(y) |\phi_0{\ra^N} \nonumber 
\end{eqnarray}
for Fig.2a and 2b where 
$T^{ij} = 2\delta^{ij} \tau^3 - \delta^{i3}\tau^j - \delta^{j3}\tau^i$,  
\begin{eqnarray}\label{piN_Fig3c} 
M_{N_1N_2}^{(e^2); ij}\bigg|_{c} &=& \frac{ie^2}{F^2} \, \cdot \, 
{^N\la}\phi_0| \int d^4x \int d^4y D_{\mu\nu}(x-y) \bar \psi_0(x) \gamma^\mu\\
&\times& T^{ij}  \psi_0(x) \bar \psi_0(y) \gamma^\nu Q 
\psi_0(y) |\phi_0{\ra^N} \nonumber 
\end{eqnarray}
for Fig.2c, 
\begin{eqnarray}\label{piN_Fig3de} 
M_{N_1N_2}^{(e^2); ij}\bigg|_{d+e} &=& - \frac{e^2}{F^2} \, \cdot \, 
{^N\la}\phi_0| \int d^4x \int d^4y D_{\mu\nu}(x-y) \bar \psi_0(x) 
\gamma^\mu\gamma^5 \\
&\times& (\varepsilon^{3ik} \varepsilon^{3jm} + 
\varepsilon^{3jk} \varepsilon^{3im})  
\tau^k G_\psi(x,y) \gamma^\nu \gamma^5 \tau^m \psi_0(y) |\phi_0{\ra^N}  
\nonumber 
\end{eqnarray}
for Fig.2d and 2e, 
\begin{eqnarray}\label{piN_Fig3f} 
M_{N_1N_2}^{(e^2); ij}\bigg|_{f} &=& \frac{ie^2}{F^2} \, \cdot \, 
{^N\la}\phi_0| 
\int d^4x \int d^4y D_{\mu\nu}(x-y) \bar \psi_0(x) \gamma^\mu\gamma^5 \\
&\times& \varepsilon^{3ik} \varepsilon^{3jm}  \tau^k \psi_0(x) \bar\psi_0(y) 
\gamma^\nu \gamma^5 \tau^m \psi_0(y) |\phi_0{\ra^N}   
\nonumber 
\end{eqnarray}
for Fig.2f. 

Truncating the quark propagator to the ground state mode the $\pi N$ 
scattering amplitude at threshold including first-order radiative 
corrections is 
\begin{eqnarray}\label{M_inv_PCQM}
\hspace*{-1.1cm} 
M_{inv}^{e^2 \pi N} &=& -\frac{1}{(4\pi)^3}\int\frac{d^3 q}{\vec{q}^{\,\, 2}} 
\bigg\{ M_{f_1}^{\pi N} \biggl[ [G_E^p(-\vec{q}^{\, 2})]^2  
- \frac{19\vec{q}^{\,\, 2}}{6m_N^2} [G_M^p(-\vec{q}^{\, 2})]^2 \\ 
& &\nonumber\\
&+& \frac{114}{25} \frac{d^2_+(\vec{q}^{\, 2})}{d^2_-(\vec{q}^{\, 2})}
 G_A^2(-\vec{q}^{\, 2}) \biggr] + M_{f_2}^{\pi N} 
\biggl[ [G_E^p(-\vec{q}^{\, 2})]^2 - \frac{5\vec{q}^{\,\, 2}}{18m_N^2} 
[G_M^p(-\vec{q}^{\, 2})]^2 \biggr] \biggr\} \nonumber\\
& &\nonumber\\
&=& - \frac{1}{8R} \frac{1}{(2\pi)^{3/2}} \biggl\{ M_{f_1}^{\pi N} 
\biggl[ \frac{41}{3} - \frac{115}{2}\kappa + \frac{953}{16} \kappa^2 \biggr] 
+ M_{f_2}^{\pi N} \, \biggl[ 1 - \frac{29}{18} \kappa + 
\frac{89}{48} \kappa^2  \biggr] \biggr\} \nonumber . 
\end{eqnarray}
where 
$$\hspace*{-.9cm} M_{f_1}^{\pi N} = - \frac{4\pi\alpha}{F^2}
\bar N\{\vec{\pi}^{\, 2} - (\pi^0)^2\} N \,\,\, \mbox{and} \,\,\, 
M_{f_2}^{\pi N} = - \frac{4\pi\alpha}{F^2}\bar N 
\{ \vec{\pi}^{\, 2} \tau_3 - (\vec{\pi} \vec{\tau})\pi^0 \}  N$$ 
and 
$$d_{\pm}(\vec{q}^{\, 2}) = 1 \pm \frac{\vec{q}^{\, 2} R^2}{4} 
\frac{\kappa}{1 - 2\kappa} .$$ 
The contribution of the Coulomb photons to the amplitude $M_{inv}^{e^2 \pi N}$ 
is pa\-ra\-me\-tri\-zed by the proton charge form factor $(G_E)$, transverse 
photons are related to the proton magnetic $(G_M)$ and axial nucleon $(G_A)$ 
f.f. where the latter is given by \cite{PCQM}  
\begin{eqnarray}
G_A(-\vec{q}^{\, 2}) = g_A \exp\biggl(-\frac{\vec{q}^{\, 2} R^2}{4}\biggr) 
d_-(\vec{q}^{\, 2}) .
\end{eqnarray}
Again, as in the case of e.m. mass shifts, the amplitude 
$M_{inv}^{e^2 \pi N}$ is gauge-independent. 
In ChPT the corresponding amplitude is given by \cite{Meissner2} 
\begin{eqnarray}\label{M_inv_ChPT}
\hspace*{-.4cm} 
M_{inv}^{e^2 \pi N}|_{ChPT} = f_1 M_{f_1}^{\pi N} + 
\frac{f_2}{4} M_{f_2}^{\pi N} . 
\end{eqnarray}
Comparing of Eqs. (\ref{M_inv_PCQM}) and (\ref{M_inv_ChPT}) we get the 
same expression for $f_2$ as already obtained from the e.m. mass 
shift (\ref{f2_estimate}). We also deduce the following relations: 
\begin{eqnarray}\label{f_1andf_2} 
f_1 &=& - \frac{1}{8R (2\pi)^{3/2}} 
\biggl[ \frac{41}{3} - \frac{115}{2} 
\kappa + \frac{953}{16} \kappa^2 \biggr] , \nonumber \\ 
f_3 &=& \frac{1}{8R (2\pi)^{3/2}} \biggl[ \frac{35}{3} - \frac{785}{18} \kappa 
+ \frac{1913}{48} \kappa^2 \biggr] . \nonumber
\end{eqnarray} 
The predicted ratio for $f_1/f_2$ depends on only one model parameter $\rho$ 
(or $\kappa$) which is related to the axial nucleon charge $g_A$ calculated 
at zeroth order. In addition, the constants $f_1$, $f_2$ and $f_3$ depend on 
the size parameter $R$ of the bound quark. For our "canonical" set of 
parameters, $g_A=1.25$ and $R=0.6 \pm 0.05$ fm, used in the calculations of 
nucleon e.m. form factors and meson-baryon sigma terms \cite{PCQM} we obtain: 
\begin{eqnarray}
& &f_1 = -19.5 \pm 1.6 \,\,\, \mbox{MeV}, \hspace*{1cm}
   f_2 = -8.7 \pm 0.7 \,\,\, \mbox{MeV}, \\ 
& &f_3 = 18 \pm 1.5 \,\,\, \mbox{MeV}, \hspace*{1.6cm} 
\frac{f_1}{f_2} = 2.2 . \nonumber 
\end{eqnarray}
Using these values of $f_1$ and $f_2$ we can estimate the isospin-breaking 
correction to the energy shift of the $\pi^- p$ atom in the $1s$ state. 
The strong energy-level shift $\epsilon_{1s}$ of the $\pi^- p$ atom is given 
by the model-independent formula \cite{piP-atom}: 
$\epsilon_{1s} = \epsilon_{1s}^{LO} +  \epsilon_{1s}^{NLO} = 
\epsilon_{1s}^{LO} (1+\delta_{\epsilon})$ where the leading order (LO) or 
isospin-symmetric contribution is $\epsilon_{1s}^{LO}$ and the next-to-leading 
order (NLO) or isospin-breaking contribution is $\epsilon_{1s}^{NLO}$. 
The quantity $\epsilon_{1s}^{LO}$ is expressed with the help of the 
well-known Deser formula \cite{Deser} in terms of the $S$-wave $\pi N$ 
scattering lengths with $\epsilon_{1s}^{LO} = - 2\alpha^3\mu_c^2 
{\cal A}_{str}$ and ${\cal A}_{str} = (2a_{\frac{1}{2}} + a_{\frac{3}{2}})/3$.
The reduced mass of the $\pi^- p$ atom is denoted by 
$\mu_c=m_p M_{\pi^+}/(m_p + M_{\pi^+})$ and 
${\cal A}_{str}=(88.4\pm 1.9)\times 10^{-3} M_{\pi^+}^{-1}$ is the strong 
(isospin-invariant) regular part of the $\pi^- p$ scattering amplitude at 
threshold \cite{PSI} (for the definitions of these quantities see Ref. 
\cite{piP-atom}). In ChPT the quantity $\delta_{\epsilon}$, the ratio of NLO 
to LO corrections, is expressed in terms of the LECs $c_1$, $f_1$ and $f_2$ 
\begin{eqnarray}\label{delta_epsilon}
\hspace*{-.8cm} 
\delta_{\epsilon} &=& \frac{\mu_c}{8\pi M_{\pi^+} F^2_\pi {\cal A}_{str}} 
[ 8c_1 (M_{\pi^+}^2 - M_{\pi^0}^2) - e^2 (4 f_1 + f_2) ] 
- 2\alpha\mu_c({\rm ln}\alpha - 1) {\cal A}_{str} . 
\end{eqnarray}
The quantity $c_1$ is the strong LEC from the ChPT Lagrangian 
\cite{Meissner1,Leutwyler1} and $F_\pi=92.4~{\rm MeV}$ is the physical 
value of the pion decay constant \cite{piP-atom}. In Ref. \cite{PCQM} we 
obtained $c_1 = - 1.16 \pm 0.1$ GeV$^{-1}$ using the PCQM approach. 
Our prediction for $c_1$ is close to the value $c_1 = - 0.9 m_N^{-1}$ deduced 
from the $\pi N$ partial wave analysis KA84 using Baryon Chiral Perturbation 
Theory \cite{Leutwyler1}. Substituting the central values for our couplings 
$f_1 = -19.5$ MeV, $f_2 = - 8.7$ MeV and $c_1 = -1. 16 $ GeV$^{-1}$ into 
Eq. (\ref{delta_epsilon}), we get $\delta_{\epsilon} = - 2.8 \cdot 10^{-2}$.  
Our estimate is comparable to a prediction based on a potential model for the 
$\pi N$ scattering \cite{PSI}: \\ $\delta_{\epsilon} = -2.1 \cdot 10^{-2}$. 
 
In conclusion, we give predictions for the $O(p^2)$ electromagnetic (e.m.) 
low-energy couplings (LECs) $f_1$, $f_2$ and $f_3$ as originally set up in 
the ChPT effective Lagrangian. The magnitude of $f_2$ and its relation to 
$f_1$ and $f_3$ are obtained from an analysis of the nucleon e.m. mass shift 
and the leading radiative corrections to the $\pi N$ scattering amplitude at 
threshold. Using our values for $f_1$ and $f_2$ we also predict the 
isospin-breaking correction to the strong energy shift of the $\pi^- p $ atom 
in the $1s$ state. Latter prediction is extremely important for the ongoing 
experiment "Pionic Hydrogen" at PSI, which intends to measure the 
ground-state shift and width of pionic hydrogen ($\pi^- p$-atom) 
at the $1\%$ level \cite{Gotta}.  

\vspace*{.3cm}
{\it Acknowledgements}. 
We thank A.~Rusetsky for useful discussions. This work was supported by 
the DFG (grant FA67/25-1) and by the DAAD-PROCOPE project.

\begin{figure}
\vspace{15cm}
\includegraphics{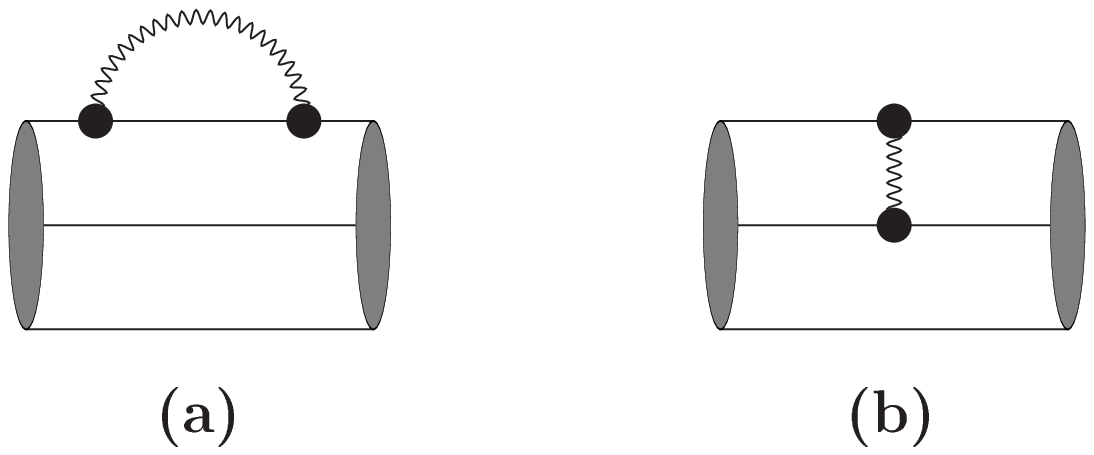}
\vspace*{-12cm}
\caption{\it Electromagnetic mass shift of the nucleon. \label{fig1} }
\end{figure}
\begin{figure}
\vspace{18cm}
\includegraphics{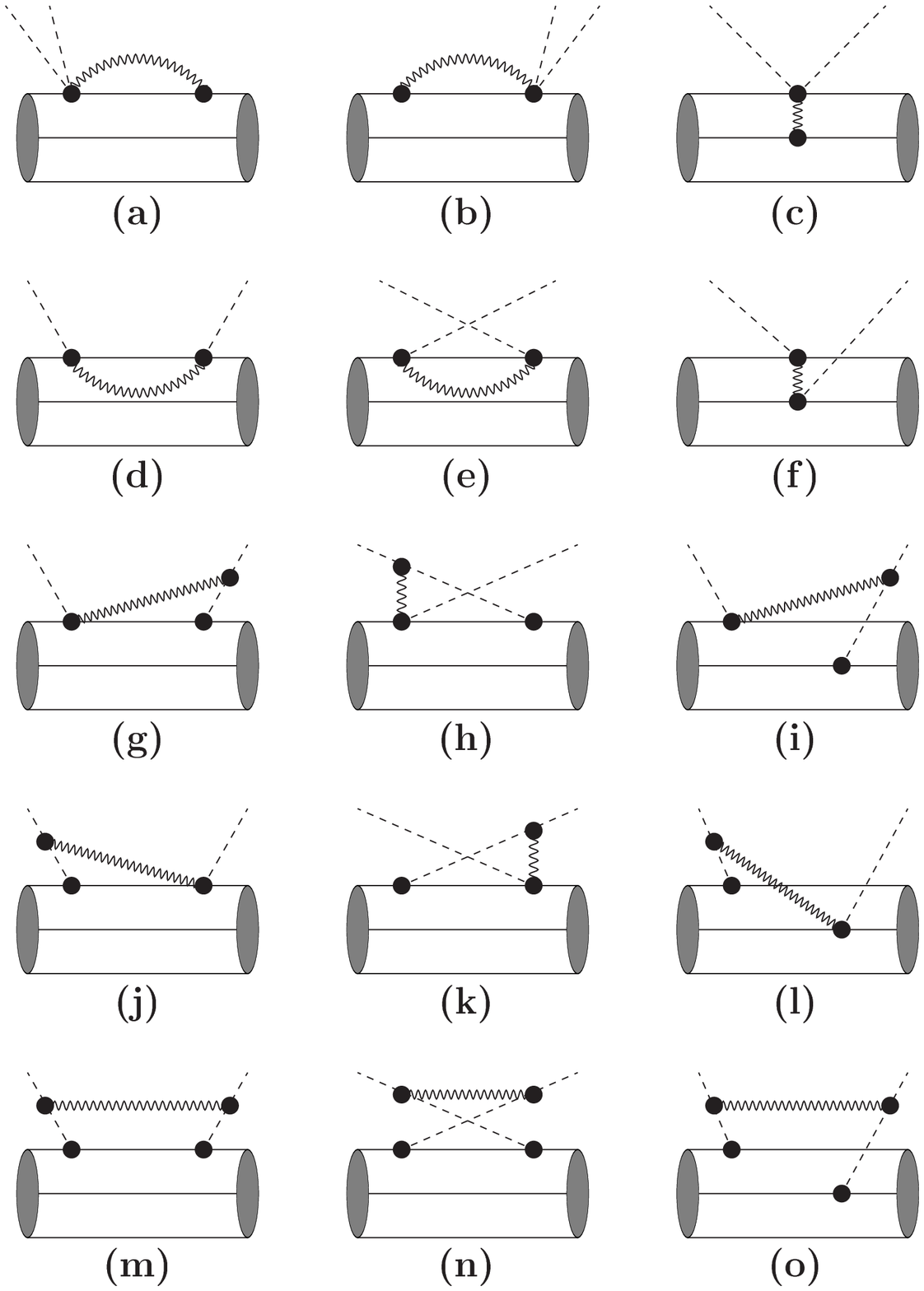}
\vspace*{-2cm}
\caption{\it Leading $e^2/F^2$ radiative corrections to the $\pi N$ 
amplitude at threshold. \label{fig2} }
\end{figure}
\end{document}